# Imaging of plasmonic modes of silver nanoparticles using high-resolution cathodoluminescence spectroscopy

*Pratik Chaturvedi[1], Keng Hsu[1], Anil Kumar[2], James C. Mabon[3] & Nicholas Fang[1*]*

1. Department of Mechanical Science & Engineering,

2. Department of Electrical and Computer Engineering,

3. Frederick Seitz Materials Research Laboratory,

University of Illinois at Urbana-Champaign, 1206 W. Green St., Urbana, IL 61801, USA

*nicfang@illinois.edu

**ABSTRACT**

Cathodoluminescence spectroscopy has been performed on silver nanoparticles in a scanning electron microscopy setup. Peaks appearing in the visible range for particles fabricated on silicon substrate are shown to arrive from excitation of out of plane eigenmodes by the electron beam. Monochromatic emission maps have been shown to resolve spatial field variation of resonant plasmon mode on length scale smaller than 25nm. Finite-difference time-domain numerical simulations are performed for both the cases of light excitation and electron excitation. The results of radiative emission under electron excitation show an excellent agreement with experiments. A complete vectorial description of induced field is given, which complements the information obtained from experiments.

**Keyword list:** Cathodoluminescence, Plasmonics, Nanoparticles, Extinction, Eigenmodes

A multitude of optical phenomena at the nanoscale are made possible by resonant surface plasmons in artificially structured metal systems. These optical phenomena often give rise to properties that are difficult to obtain in natural materials. An entire new generation of artificial materials in the emerging field of plasmonics is designed to harness these properties through nanoscale engineering. These materials find tremendous applications in chemical and biological sensing.[1, 2] By simple surface patterning a thin metal film, it is possible to engineer its surface modes over a wide range of frequency.[3] Highly localized optical modes associated with patterned surfaces with nanoscale features (<~200nm) and the sensitivity of these modes to local refractive index finds tremendous potential in realizing compatible and efficient sensors. These optical modes known as localized surface plasmon resonance (LSPR) modes are responsible for producing strong scattering and extinction spectra in metal nanoparticles such as silver and gold. Exploiting local electromagnetic field enhancement associated with these plasmonic structures has led to several interesting applications such as enhanced fluorescence,[4] enhanced photo-carrier generation[5] and other nonlinear effects such as second harmonic[6] and high-harmonic generation[7]. Often the field is confined spatially on length scales on the order of 10-50nm and varies strongly with particle shape, size and material composition.[8] Unfortunately, diffraction-limited optical imaging techniques do not have enough spatial resolution to image these plasmon modes or precisely locate the "hot-spots" responsible for producing enormous enhancement. Near-field scanning optical microscopy (NSOM) has been used to investigate these plasmon modes,[9] however, the resolution is limited by the tip size (~50-100nm). On the other hand, electron beam based characterization techniques such as cathodoluminescence (CL) and electron energy loss spectroscopy (EELS) are able to excite and image plasmon modes with very high spatial resolution. EELS for example has been demonstrated to resolve plasmon modes on length scale below 18nm.[10] EELS technique, however, has to be performed in a transmission electron microscope (TEM), where it detects the inelastically scattered electrons and the loss suffered by electron beam in exciting surface plasmons. Although the technique has been described as one with the best spatial and energy resolution,[10] it requires samples to be electron transparent (typically <100nm). Specialized sample preparation procedure (used for TEM) and instrumentation makes it an expensive alternative and infeasible for samples on thick substrate.

On the other hand, scanning electron microscopy (SEM) based CL technique does not suffer from this limitation. CL (in both SEM and TEM mode) has been utilized to image plasmon modes of particles and antennas of various shapes.[11-14]

CL has been used in materials science as an advanced technique for examination of intrinsic structures of semiconductors such as quantum wells[15, 16] or quantum dots[17, 18]. Typically, a tightly focused beam of electrons impinges on a sample and induces it to emit light from a localized area down to 10-20 nanometers in size. By scanning the electron beam in an X-Y pattern and measuring the wavelength and intensity of light emitted with the focused electron beam at each point, a high resolution map of the optical activity of the specimen can be obtained. In traditional cathodoluminescence of semiconductors, impingement of a high energy electron beam will result in the excitation of valence electrons into the conduction band, leaving behind a hole. The detected photon emission is actually a result of electron-hole recombination process. In the case of metallic nanostructures however, the photons are produced as a result of excited plasmons, i.e. collective motion of the conduction electrons induced by the fast moving electrons, and these induced charges can act back on the electron beam, causing it to lose energy as detected in EELS. In CL spectroscopy, we are able to detect radiation due to the oscillating plasmon on metallic structures, allowing quantitative study of the local field (Figure 1). Mechanism of this radiation has recently been presented[13, 19]. While, photon emission from semiconductor materials on interaction with electron beam is well understood, CL from plasmonic nanostructures is a relatively new field and deserves more attention.

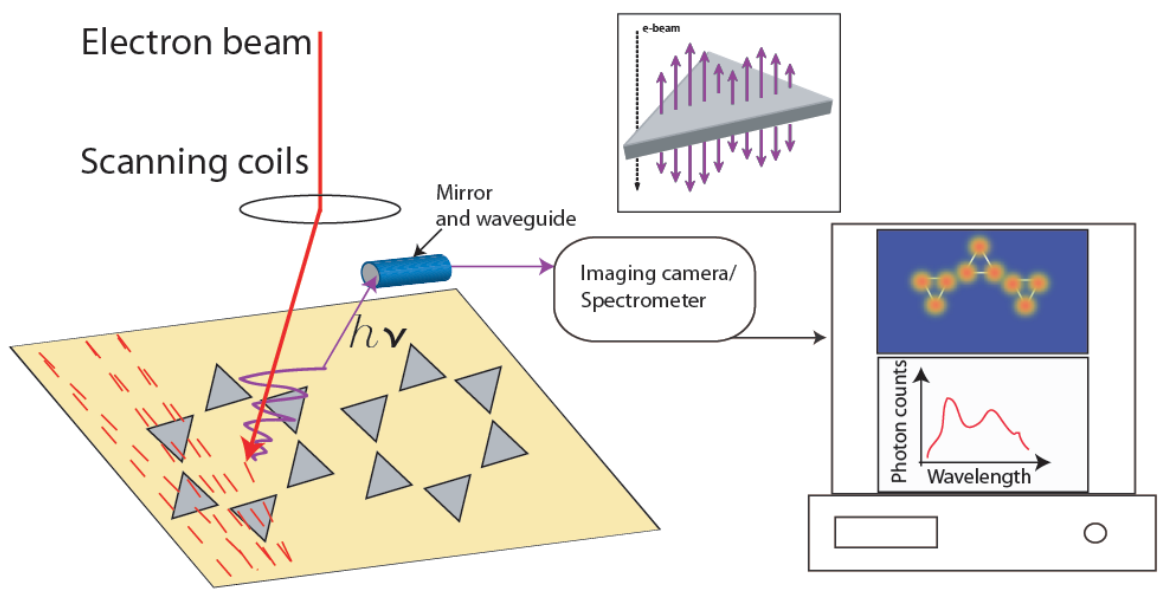

**Figure 1:** Schematic illustration of CL spectroscopy and imaging technique performed in scanning electron microscope. *Inset:* Passing electron beam induces current/electromagnetic oscillations in a metallic particle. These oscillations known as surface plasmon modes are responsible for radiation detected in CL.

In this study, we investigate the plasmon modes of silver (Ag) triangular nanoparticles using CL imaging and spectroscopy. The study is intended to identify the local modes of these nanoparticles. While it is understood that the excitation of plasmons in these metallic nanoparticles is responsible for the field enhancement effect, it is a challenge to identify the local fields associated with these plasmons. Several theoretical studies have identified the plasmon eigenmodes of triangular nanoparticles,[20-22] only a few experimental studies have demonstrated a resolution capability of mapping the spatial field variation associated with these plasmon modes.[10, 23, 24] In this work, we report direct excitation and emission of decoupled surface plasmon modes with CL spectroscopy (in SEM chamber) on triangular nanoparticles. In spectroscopic mode with monochromatic photon maps, we are able to distinguish the dramatic spatial variation of resonant plasmon mode on length scales smaller than 25nm. Numerical simulations were performed to identify the plasmon eigenmodes of triangular particles using a commercial finite-difference time-domain (FDTD) simulator.[25] Both electron beam excitation and a more conventional plane wave scattering type calculations are performed to stress the differences between light excitation and electron excitation. Electron excitation calculations are performed by modeling the moving electron charge as a series of closely spaced dipoles with temporal phase delay governed by the velocity of electron. We also incorporate substrate effect into our calculations. We illustrate that while normally incident light can excite in-plane eigenmodes, electron beam is capable of exciting out of plane dipole mode of the particles. Under resonance conditions CL emission maps depict a standing-plasmon mode pattern with a spatial full-width at half-maximum (FWHM) of 35nm at 400nm wavelength. This is in excellent agreement with simulations which show a FWHM of 30nm.

**Results and discussion**

Conventionally, nanoparticles are characterized by their extinction spectra. The peaks observed in absorption or scattering spectra of particles under light excitation reveal resonant wavelengths of certain plasmon eigenmodes of the particle. While light excitation can couple to low frequency plasmon eigenmodes, it is hard to excite high-frequency plasmon states due to large momentum mismatch.[26] Electron excitation on the other hand can couple to high-frequency plasmon modes and recently it has been described to directly reveal the local density of plasmon states.[27] While optical techniques are limited in their resolution capability to image the plasmon eigenmodes, electron excitation on the other hand is potentially capable of resolving details below 10s of nanometers. Resolving surface plasmon modes and understanding the underlying physics is crucial to design better plasmonic devices tailored to specific applications.

For the purpose of this study we fabricated 40nm thick Ag equilateral triangular nanoparticles with ~200nm edge length arranged in a hexagonal lattice. These particles are fabricated on silicon substrate and the shortest distance between two adjacent particles is >100nm. Silicon is chosen as the substrate material, to suppress background cathodoluminescence in the wavelength range of interest (near-UV and visible). For the purpose of numerical simulations we model and analyze single nanoparticle. This is because experimentally the interaction distance between electron beam and the particle is limited to few 10's of nms and hence, the excitation of plasmon modes is insensitive to particle coupling over ~100nm spacing. This is especially true for particles on a non-plasmonic substrate such as silicon. We have performed spectrally resolved CL imaging experiments on these triangular nanoparticles on Si substrate. Emission spectrum of the particle induced by the electron beam passing through nearby the tip of the particle reveals a resonance peak at 405nm and a secondary peak at 376nm (Figure 2, blue). This is in excellent agreement with simulations which indicate a resonance peak at 430nm and secondary peak at 385nm under tip excitation (Figure 2, red). It is to be noted that in this simulation, Si substrate has been approximated as non-dispersive loss-less material with an average refractive index of 4.8 (see methods).

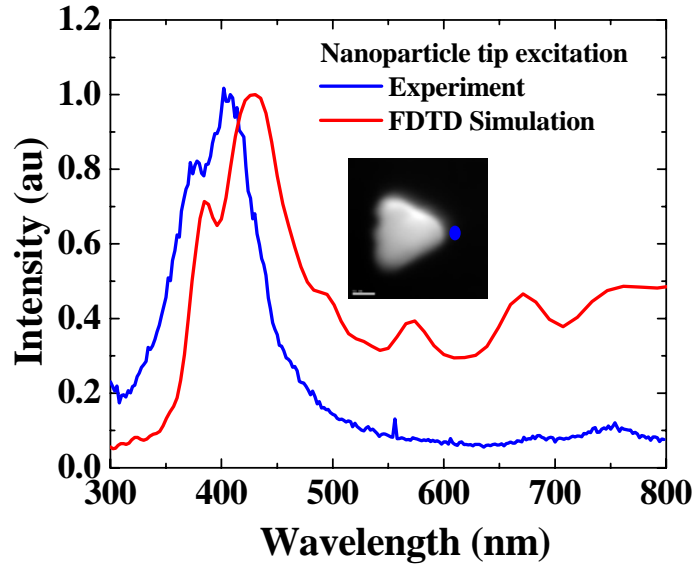

**Figure 2:** Luminescence spectrum collected from triangular nanoparticle under tip excitation (blue). The spectrum was corrected for grating response function. Corresponding simulated radiation spectrum (red). *Inset:* SEM image of the particle with blue dot showing the position of the electron beam. The scale bar is 50nm.

Our experimental setup consists of a paraboloidal mirror, which is placed between the sample stage and the electron beam in a SEM chamber. The electron beam passes through an aperture in the mirror to the sample surface. The sample is at the focus of the mirror which lies 1mm below it. Light emitted by the sample is collected by the mirror and is directed to the detectors through a light guide. Spectrally resolved measurements are performed using a monochromator (Czerny-Turner type). Light passing through a monochromator allows taking a spectrum, as well as images at a selected wavelength. In panchromatic mode of imaging, light skips the monochromator and all of the light is carried to the detection optics. The measurements are performed using a 15kV electron beam and a photo multiplier tube (PMT) detector with sensitivity encompassing near-ultraviolet (UV) and visible wavelengths (250-850nm).

Figure 3a is the secondary electron image (SEI) of triangular nanoparticle which gives the topographic information about the specimen. Figure 3b is a panchromatic CL image (PanCL). In panchromatic mode all of the emitted light is collected by the detector and hence the intensity at each pixel represents the integrated photon counts in the sensitivity range of the detector. PanCL image clearly depicts plasmon induced luminescence in Ag nanoparticle. This luminescence arises due to induced electromagnetic field on the nanoparticle caused by the external field of incoming electrons. The way this image is acquired is similar to SEM mapping i.e. by raster scanning the electron beam and collecting emitted photons rather than secondary electrons as done in scanning electron imaging. The collected light when passed through a grating monochromator allows resolving spectral features as shown in figure 2. As an experimental reference, we have also recorded the emission from flat silver film which reveals a sharp bulk plasmon peak at 325nm and surface plasmon peak at 340nm (Figure 3c, blue). The location of these resonant peaks on flat silver matches with the material permittivity data[28] within ±5nm. The nature of these peaks (bulk vs surface) was further confirmed by a separate CL experiment, where we coated the flat silver film with ~5.5nm thick alumina ($Al_2O_3$) coating using Atomic Layer Deposition (ALD). In this case, we observe a sharp peak at 325nm and a relatively broad peak at 357nm (Figure 3c, red). This confirms that the peak at 325nm corresponds to bulk plasmon of silver; whereas the second peak at 340nm (357nm) corresponds to surface plasmons at silver-air (silver-alumina) interface.

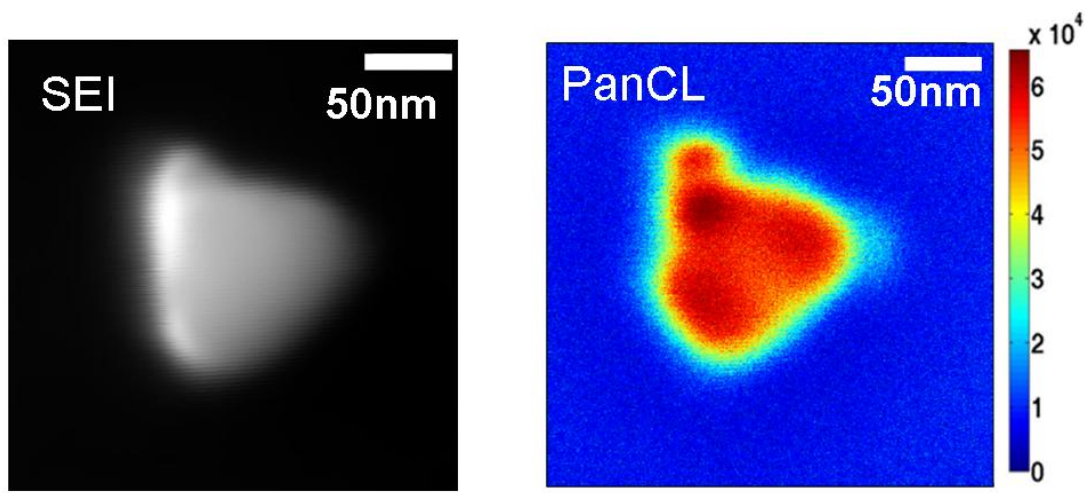

(a)                          (b)

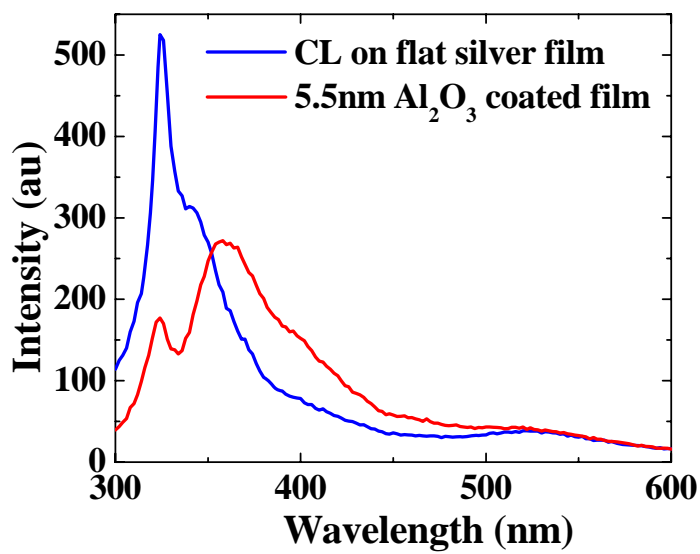

(c)

**Figure 3:** (a) Scanning electron micrograph of triangular nanoparticle. (b) Panchromatic CL image of the same. (c) Luminescence spectrum collected from a flat silver film.

Apart from emission spectra, monochromatic photon emission maps are acquired. These monochromatic CL images were obtained by setting the grating monochromator to a specific wavelength and scanning the electron beam over the nanoparticle. These emission maps acquired by raster scanning the electron beam reveal the standing-wave patterns of surface plasmons.[10, 11] These standing-wave patterns are observed only under resonance conditions, i.e. when the field produced by the electron beam couples strongly to eigenmodes of the particle. Monochromatic CL image obtained at 400nm wavelength (with bandwidth 5.4nm and 5ms dwell time at each pixel) shows strong luminescent intensity when the electron beam scans over the tip region of the particle (Figure 4a). An image obtained at a wavelength of 355nm (Figure 4b) depicts no discernible features in spatial variation of emission suggesting non-resonant excitation.

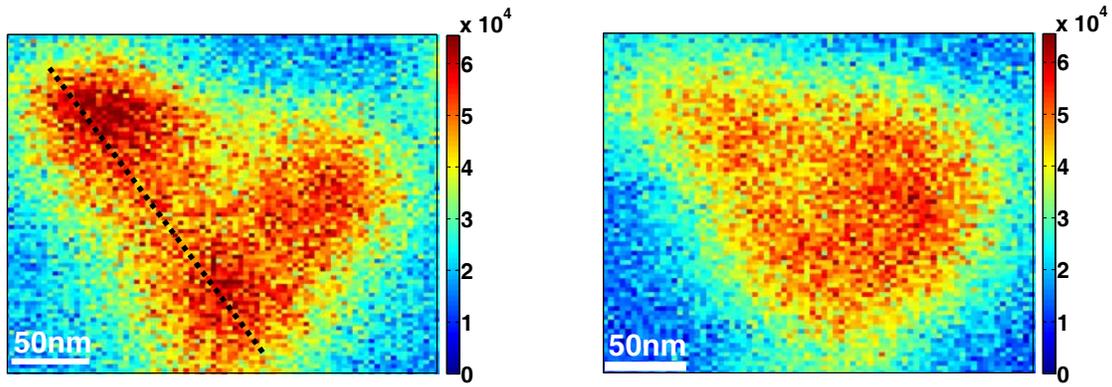

(a)          (b)

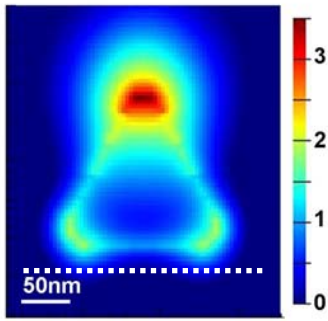 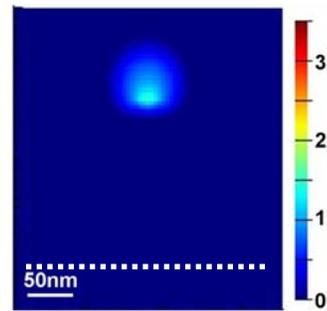

(c)          (d)

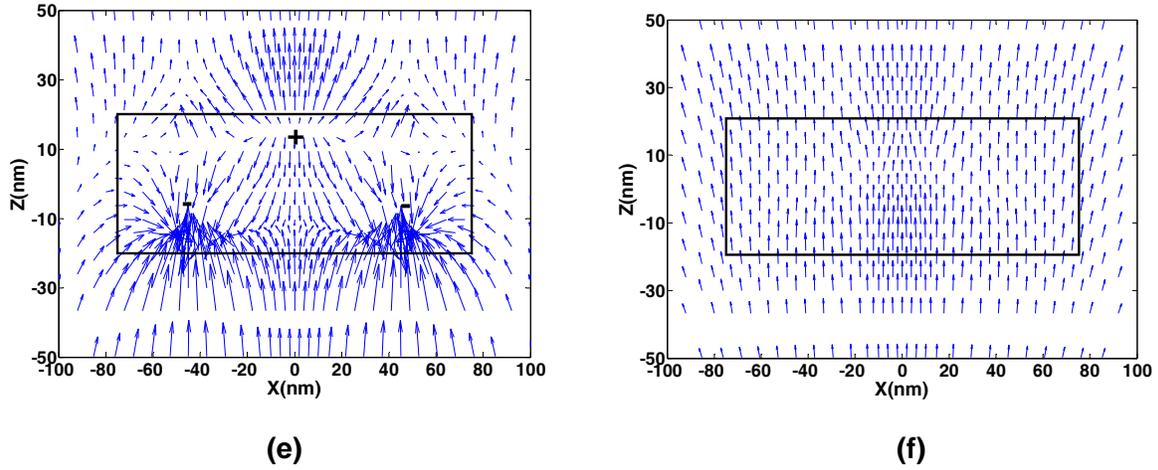

**(e)**  **(f)**

**Figure 4:** (a) Monochromatic photon emission map acquired at 400nm wavelength and (b) 355nm wavelength. (c) Simulated electric field intensity with tip excitation at 400nm wavelength for triangular nanoparticle on substrate (n = 4.8). (d) Intensity at 355nm wavelength. The color scale is in arbitrary units in log scale for (c) and (d). (e) Simulated vector plot of electric field at 400nm wavelength showing out of plane dipole mode excitation near the tip regions of the particle. The location of the plane is indicated by white dotted line in (c). Electron beam travels in z direction and the particle boundary is shown by black lines. (f) Vector plot at off-resonance wavelength of 355nm.

**Numerical simulations:**

While CL experiments are limited to mapping the emitted light intensity by scanning the electron beam, numerical simulations allow us to map the field with fixed electron beam position. It is observed that under resonance conditions the induced electromagnetic field from the electron beam extends across the entire nanoparticle. This is illustrated in figure 4c where the electron beam is located near the topmost tip of the particle and the intensity is plotted at 400nm wavelength. Notice the strong intensity near the tips of the particle, in contrast away from resonance (λ = 355nm), the induced field is weak and localized near the probe position (figure 4d). Hence, the monochromatic emission maps acquired by scanning the electron beam under resonance condition illustrate strong luminescence near the tip regions of the particle and no spatial variation (above the noise level) away from resonance.

It should be noted that the emission pattern obtained at 400nm wavelength is very similar to the in-plane tip eigenmode of the triangular particle illustrated in earlier theoretical[20, 21] and experimental[10, 23, 24] studies. However, given the dimensions of the particle, the in-plane tip eigenmode occurs at much longer wavelengths. Our simulations indicate that the resonance at 400nm wavelength corresponds to out of plane dipole mode excitation by the electron beam. This is in strong contrast to light excitation, where a normally incident plane-wave excites electromagnetic field that correspond to in-plane charge oscillations. An electron beam on the other hand can excite out of plane charge oscillations. This is illustrated in figure 4e which plots the simulated vector distribution of electric field in a plane parallel to the direction of electron beam. Under non-resonance condition (λ = 355nm), the induced field does not show charge oscillations (figure 4f).

*Light excitation:*

To further illustrate the differences between light excitation and electron excitation, we have calculated the scattering properties of triangular nanoparticle under plane-wave illumination. When light is resonantly coupled to the plasmon modes of a nanoparticle, it leads to strong scattering and absorption of the incident field. Thus, the resonance modes can be identified based on extinction spectrum of the particle. Figure 5a presents the extinction spectra of isolated equilateral triangular nanoparticles (200nm edge length, 40nm thickness) suspended in air. We observe a dipolar plasmon peak at 677 nm and quadrupole peak at 400nm (black curve). The polarity of these peaks is identified based on vectorial description of the polarization response of the particle. These resonant peaks represent the well known in-plane "tip" (dipole) and "edge" (quadrupole) eigenmodes of the triangular particle (plane here refers to the plane of the particle).[10, 20, 29] It is to be noted that this result is for idealistic nanoparticle geometry with sharp tips. Deviation from this geometry such as rounding of tips is known to cause blue-shift of plasmon resonance peaks.[20] This happens because of effective reduction in volume of the particle that leads to reduction in its polarizability. Figure 5a and 5b illustrate this effect. With a tip radius ($r$) of 20nm on all corners a blue-shift of 68nm is observed in dipolar resonance. It is observed that the shift in quadrupole mode is not significant. Rounding of tips also reduces the maximum field enhancement factor which usually is localized to the corners of the tips. Hence, the peak extinction efficiency also reduces with increasing radius of curvature of tips. It is worth mentioning here that for a particle with 3-fold rotational symmetry, the extinction cross-section under plane wave excitation with two orthogonal polarizations (parallel and perpendicular to the edge of the triangle) are identical at normal incidence. However, the charge oscillations are oriented along the direction of incident polarization.

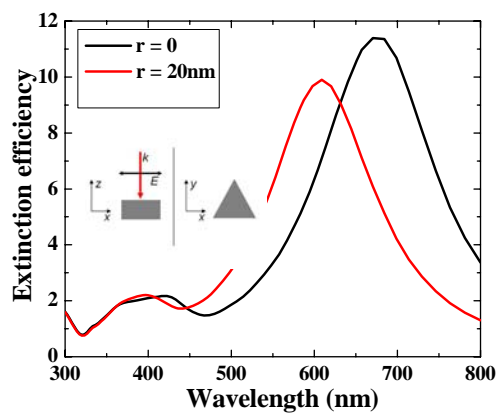
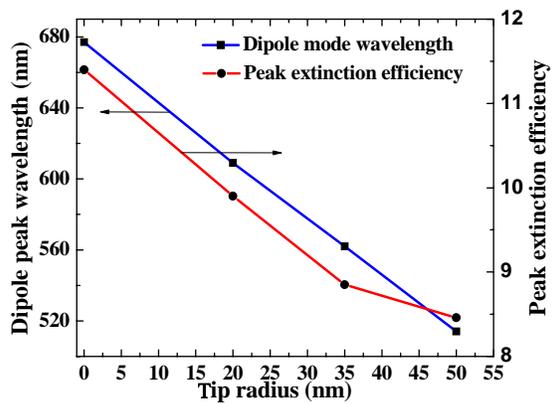

(a)

(b)

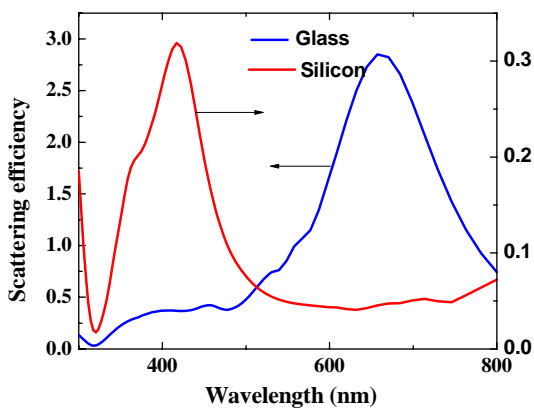
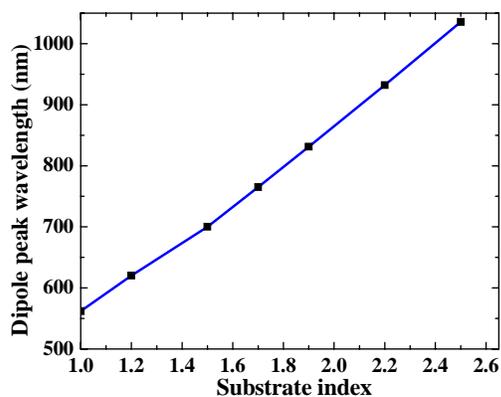

(c)

(d)

**Figure 5:** **(a)** Extinction spectra of triangular nanoparticle and effect of rounding of tips. *Inset*: Illumination and polarization direction of plane wave. **(b)** Shift of dipolar resonance and reduction in extinction efficiency with rounding of tips. **(c), (d)** Effect of substrate in plasmon resonances of the particle.

Next, we consider the effect of substrate. Scattered field characteristics of a particle vary significantly on interaction with a non-homogeneous dielectric environment such as in the presence of a substrate. Figure 5c illustrates this effect where the scattering cross-section of triangular particles situated on top of a substrate is plotted. In accordance with experimental setups suitable for light scattering measurements by particles on a substrate, we incorporate light incidence from the top (air side) and only the scattered power in top half plane is considered in the calculations. It is observed that the dipolar resonance of the particle on a substrate is red-shifted with respect to its free-space resonance (figure 5d). Qualitatively, the red-shift of the resonance can be explained by the increase in the effective permittivity of the surroundings. Particles appear larger with respect to the effective illumination wavelength in high-index surroundings which causes an increased retardation effect. Thus, for a non-dispersive substrate the amount of red-shift can be roughly approximated by taking the average refractive index of the surroundings. It is worth mentioning that the shift in higher order resonances such as quadrupole is not as significant as dipolar mode. Hence, for a high-index substrate such as silicon only in-plane quadrupolar resonance is observed (with light excitation) in visible range for the particle size under consideration.

*Electron excitation:*

As mentioned before scattering or extinction spectrum of a particle under plane wave excitation can be significantly different than its CL or EELS spectrum. While a plane wave represents a volumetric excitation source; on the other hand a highly focused electron beam represents a localized probe which gives information about local density of plasmon states.[27] Furthermore, the two electron characterization techniques also probe different properties of the particle; while EELS measures the total energy loss suffered by the electron in inducing electromagnetic fields on the particle, CL measures only part of the induced field which is radiated out. The two spectra EELS and CL would coincide if there were no losses in the system and the entire induced field is radiated. To numerically investigate the radiative modes that can be excited by a fast moving electron in CL setup, the electron beam can be modeled as a line current density source. The current density due to a moving electron can be written as: $\vec{J}(\vec{r},t) = -ev\hat{z}\delta(z-vt)\delta(x-x_0)\delta(y-y_0)$, where $e$ represents electronic charge, $v$ stands for velocity of electron, $x_0$ and $y_0$ represent the position of the electron beam and $z$ is the direction of electron travel. In FDTD simulation approach, this current density due to a moving charge can be modeled as a series of dipoles with temporal phase delay that is governed by electron velocity (see methods). The radiative energy component of the induced electromagnetic field is calculated by integrating the Poynting vector normal to an arbitrary large surface in the upper half-plane. Figure 6a presents the radiation spectra of triangular nanoparticle in free-space on excitation with a moving electron charge. Because of the inherent anisotropy of the particle, we model two distinguished cases 1) when the electron beam is close to the tip of the particle 2) when it is close to an edge of the particle. It is found that for both of these two cases main resonance occurring at ~600nm range (613nm for tip and 622nm for edge excitation) correspond to in-plane dipolar mode of the nanoparticle, as illustrated in figure 6b. However, the weaker resonance occurring at ~380nm corresponds to out of plane dipole mode. Since the thickness of the particle is much smaller than its edge length, out of plane dipole resonance occurs at much shorter wavelength compared to in-plane resonance. From the spectra it is evident that tip excitation is more efficient in exciting in-plane dipolar mode compared to edge excitation. Moreover, when the electron beam is close to the edge of the particle it also excites in-plane quadrupole mode at 400nm wavelength, which leads to broadening of the peak observed in the short wavelength range.

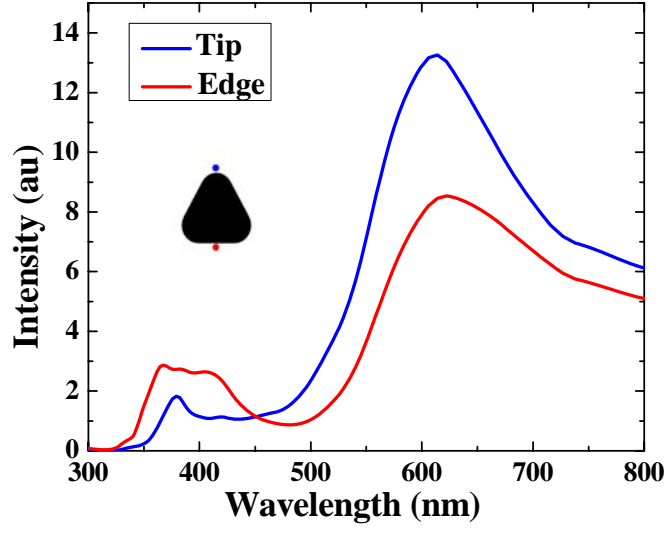

(a)

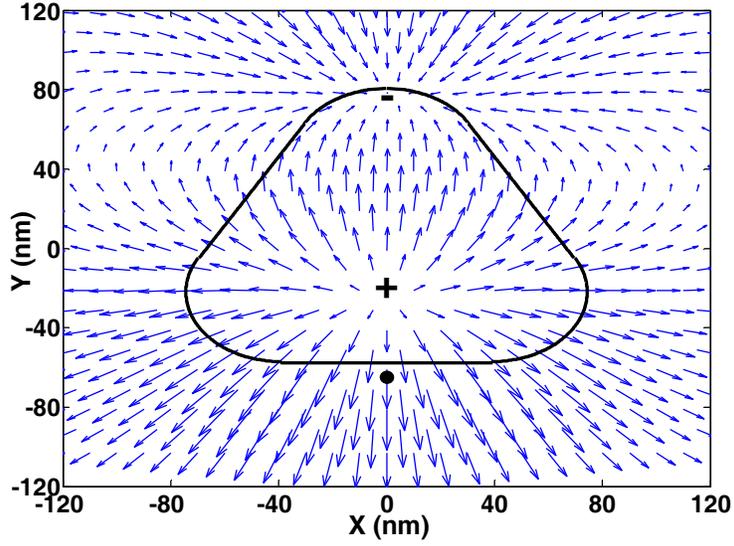

(b)

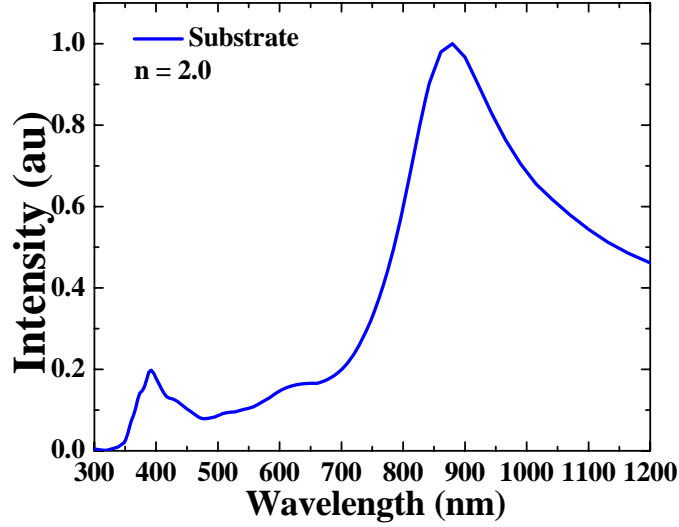

**(c)**

**Figure 6: (a)** Simulated radiation spectra of triangular nanoparticle in free-space upon excitation with electron beam. *Inset*: Position of the electron beam for tip (blue) and edge (red) excitation cases. **(b)** Vector plot of electric field for edge excitation case at 622nm wavelength, 10nm away from particle surface showing the excitation of in-plane dipole mode. The position of the electron beam is marked by black dot. **(c)** Effect of substrate on particle resonance upon excitation with electron beam (tip excitation). The substrate is assumed to be of constant refractive index *n* = 2.

We can extend some key observations from plane wave simulations to understand the effect of substrate on particle resonance under electron excitation. The in-plane dipole mode should red-shift because of increase in the index of surroundings. This is indeed observed from simulations (Figure 6c). The shift in out of plane dipolar mode is not as significant. This is expected since the thickness of the particle is small, the out of plane modes experience lower retardation effects.

This suggests that for these triangular nanoparticles fabricated on high-index substrates such as silicon, electron beam can predominantly excite out of plane dipole mode and in-plane quadrupole mode in visible wavelength range. This is indeed observed in our CL experiments and simulations. The experimentally observed spectrum shows some minor differences compared to simulations (figure 2). This may be attributed to the approximations we have made in our simulations. In electron excitation case, we did not include the dispersive properties of silicon substrate in our simulations. Secondly, in simulations the radiation spectra consists of photons integrated over the entire top half space. In our experiments, the collection angle of the mirror is limited to a cone angle of 160 degrees. Under the light of these differences, the experimental spectrum is in good agreement with simulations.

**Resolution:**

It is evident that CL technique allows high-resolution mapping of plasmon modes. The quantification of the optical resolution of the technique deserves special attention. While, it may seem that the resolution of the technique would ultimately be limited only by electron-beam diameter, as in the case of secondary electron images, however, this may not be the case. This is because in secondary electron imaging, the incoming electron beam knocks off low energy secondary electrons (<50eV); the physical nature of this process allows high-resolution topographic image acquisition (1-5nm). However in CL imaging, photon emission can occur even when the electron beam is at a distance away from the particle. Electron beam can induce luminescence in a structure without physically passing through it, as indicated by our simulations. As a rough estimate, for 15keV electron beam, this interaction length can be as large as 18nm for light emitted at 400nm wavelength.[13] Full-width at half-maximum (FWHM) of plasmon eigenmode can be taken as a criterion to estimate resolution.[10] By fitting a Gaussian to the eigenmode at 400nm wavelength near the tip region of the particle, we observe a FWHM of 35nm experimentally. In comparison, simulations show a FWHM of 30nm. A clear spatial modulation of the eigenmode above the noise level is detectable on length scale as short as 25nm as illustrated in figure 7, which plots the variation of radiation intensity along the edge of the particle as marked in figure 4a.

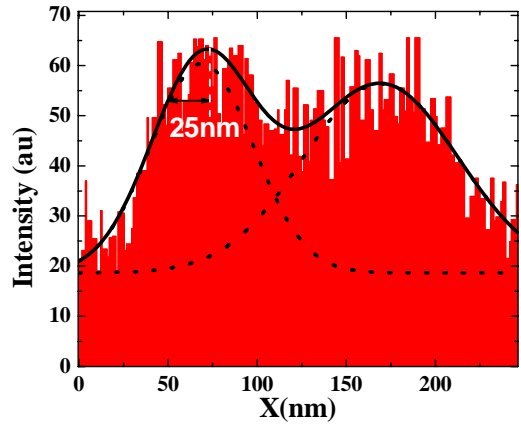

**Figure 7:** Variation of cathodoluminescence emission along the edge of the particle (marked in figure 4a) at 400nm wavelength.

**Conclusions**

In this work CL imaging technique was utilized to image plasmon modes of Ag triangular nanoparticles with high-spatial (~25nm) and spectral resolution (~5.4nm). Spectroscopic analysis when combined with monochromatic imaging helps us to identify different channels of emission of plasmon modes. The process of radiative emission of plasmon modes in CL setup was modeled using FDTD approach. Simulations indicate that in contrast to light excitation, electron beam not only excites the in-plane eigenmodes of nanoparticles but is also able to excite out of plane modes. Because of the inherent anisotropy of the triangular particle, the position of the electron beam also influences the excitation of eigenmodes. This was presented in the context of "tip" and "edge" excitation of the particle. These results provide a better understanding of excitation and imaging of plasmon modes using CL spectroscopy.

**Methods:**

*Fabrication:*

For the purpose of this study, Ag nanostructures were fabricated on silicon substrate. The samples were fabricated using a novel solid-state superionic stamping (S4) process.[30, 31] This process utilizes a pre-patterned stamp made of a superionic conductor such as silver sulfide which supports a mobile cation (silver). The stamp is brought into contact with a substrate coated with a thin silver film. On the application of an electrical bias with the substrate as anode and a metallic electrode at the back of the stamp as cathode, a solid state electrochemical reaction takes place only at the actual contact at the interface. This reaction progressively removes a metallic layer of the substrate at the contact area with the stamp. Assisted by a nominal pressure to maintain electrical contact, the stamp gradually progresses into the substrate, generating a pattern in the silver film complementary to the pre-patterned features on the stamp. Silver sulfide stamps were patterned using focused ion beam technique. A very thin (~2nm) chromium (Cr) layer is used as the adhesion layer for silver film on silicon. The fabricated structures are coated conformally with very thin dielectric layer (anatase $TiO_2$, 5 monolayer ~ 2.5Å) using atomic layer deposition (ALD) to protect the samples from environmental and electron beam damage. Excellent pattern transfer fidelity of the S4 approach down to sub-50nm resolution and ambient operating conditions make this process suitable for low-cost, high-throughput patterning of plasmonic nanostructures, such as presented in this study.

*Simulation:*

Light excitation: To numerically compute absorption and scattering by triangular nanoparticles, we utilize the total-field scattered-field (TFSF) formulation with FDTD approach. In this approach, the computation region is divided into two sections – one where the total field (incident + scattered) is computed and the second where only scattered field is computed. The particles are excited by a normal incident plane wave. Absorption and scattering cross-section are computed by monitoring the net power inflow in the total-field region near the particle and net power outflow in the scattered field region, respectively. Extinction cross-section is the sum of absorption and scattering cross-section of the particle. Our numerical calculations suggest that for Ag nanoparticles scattering is approximately an order of magnitude larger than absorption. The material properties used in the calculation are obtained from generalized multi-coefficient model[25] that fits the dispersion data obtained from Palik[32]. This approach is more accurate for broadband simulations than fitting a single material model such as Drude or Lorentz.

Electron excitation: The electron beam has been modeled as a series of closely spaced dipoles each with temporal phase delay according to the velocity of the electron beam. In the absence of any structure, electron beam moving at a constant velocity does not generate any radiation. In FDTD, however, we simulate only a finite portion of the electron path and the sudden appearance and disappearance of the electron will generate radiation. To solve this problem, we run a second, reference simulation where all the structures are removed, and we can calculate the electromagnetic fields at any wavelength by taking the difference in fields between the simulations.[25] To get an accurate difference, we force the simulation mesh to be exactly the same with and without the structure. Currently, the methodology doesn't permit electron beam to pass through a lossy or dispersive substrate material. Further work in this direction is currently underway.

**Acknowledgement:**

The authors are thankful to post doctoral researchers Kin Hung Fung and Xu Jun of University of Illinois for several informal discussions. The authors also acknowledge the technical support and assistance received from James Pond and Chris Kopetski of Lumerical Solutions Inc. Cathodoluminescence experiments were carried out in the Frederick Seitz Materials Research Laboratory Central Facilities, University of Illinois, which are partially supported by the U.S. Department of Energy under grants DE-FG02-07ER46453 and DE-FG02-07ER46471. The authors are grateful for the financial support from the Defense Advanced Research Projects Agency (grant HR0011-05-3-0002), Office of Naval Research (grant N00173-07-G013) and National Science Foundation (grant CMMI-0709023).

**References:**

1. Haes, A. J.; Van Duyne, R. P., A nanoscale optical biosensor: Sensitivity and selectivity of an approach based on the localized surface plasmon resonance spectroscopy of triangular silver nanoparticles. *J. Am. Chem. Soc.* 2002, *124,* 10596-10604.

2. McFarland, A. D.; Van Duyne, R. P., Single silver nanoparticles as real-time optical sensors with zeptomole sensitivity. *Nano Lett.* 2003, *3,* 1057-1062.

3. Pendry, J. B.; Martin-Moreno, L.; Garcia-Vidal, F. J., Mimicking surface plasmons with structured surfaces. *Science* 2004, *305,* 847-848.

4. Bharadwaj, P.; Novotny, L., Spectral dependence of single molecule fluorescence enhancement. *Optics Express* 2007, *15,* 14266-14274.

5. Ishi, T.; Fujikata, J.; Makita, K.; Baba, T.; Ohashi, K., Si nano-photodiode with a surface plasmon antenna. *Japanese Journal of Applied Physics Part 2-Letters & Express Letters* 2005, *44,* L364-L366.

6. Eftekhari, F.; Gordon, R., Enhanced Second Harmonic Generation From Noncentrosymmetric Nanohole Arrays in a Gold Film. *IEEE J. Sel. Top. Quantum Electron.* 2008, *14,* 1552-1558.

7. Kim, S.; Jin, J. H.; Kim, Y. J.; Park, I. Y.; Kim, Y.; Kim, S. W., High-harmonic generation by resonant plasmon field enhancement. *Nature* 2008, *453,* 757-760.

8. Haes, A. J.; Haynes, C. L.; McFarland, A. D.; Schatz, G. C.; Van Duyne, R. R.; Zou, S. L., Plasmonic materials for surface-enhanced sensing and spectroscopy. *MRS Bull.* 2005, *30,* 368-375.

9. Yin, L.; Vlasko-Vlasov, V. K.; Rydh, A.; Pearson, J.; Welp, U.; Chang, S. H.; Gray, S. K.; Schatz, G. C.; Brown, D. B.; Kimball, C. W., Surface plasmons at single nanoholes in Au films. *Appl. Phys. Lett.* 2004, *85,* 467-469.

10. Nelayah, J.; Kociak, M.; Stephan, O.; de Abajo, F. J. G.; Tence, M.; Henrard, L.; Taverna, D.; Pastoriza-Santos, I.; Liz-Marzan, L. M.; Colliex, C., Mapping surface plasmons on a single metallic nanoparticle. *Nature Physics* 2007, *3,* 348-353.

11. Gomez-Medina, R.; Yamamoto, N.; Nakano, M.; Abajo, F. J. G., Mapping plasmons in nanoantennas via cathodoluminescence. *New J. Phys.* 2008, *10,* 105009.

12. Hofmann, C. E.; Vesseur, E. J. R.; Sweatlock, L. A.; Lezec, H. J.; Garcia de Abajo, F. J.; Polman, A.; Atwater, H. A., Plasmonic modes of annular nanoresonators imaged by spectrally resolved cathodoluminescence. *Nano Lett.* 2007, *7,* 3612-3617.

13. Yamamoto, N.; Araya, K.; de Abajo, F. J. G., Photon emission from silver particles induced by a high-energy electron beam. *Phys. Rev. B* 2001, *64,* 205419.

14. Yamamoto, N.; Nakano, M.; Suzuki, T., Light emission by surface plasmons on nanostructures of metal surfaces induced by high-energy electron beams. *Surf. Interface Anal.* 2006, *38,* 1725-1730.

15. Cibert, J.; Petroff, P. M.; Dolan, G. J.; Pearton, S. J.; Gossard, A. C.; English, J. H., Optically detected carrier confinement to one and zero dimension in GaAs quantum well wires and boxes. *Appl. Phys. Lett.* 1986, *49,* 1275-1277.

16. Chichibu, S.; Wada, K.; Nakamura, S., Spatially resolved cathodoluminescence spectra of InGaN quantum wells. *Appl. Phys. Lett.* 1997, *71,* 2346-2348.

17. Leon, R.; Petroff, P. M.; Leonard, D.; Fafard, S., Spatially Resolved Visible Luminescence of Self-Assembled Semiconductor Quantum Dots. *Science* 1995, *267,* 1966-1968.

18. Rodriguez-Viejo, J.; Jensen, K. F.; Mattoussi, H.; Michel, J.; Dabbousi, B. O.; Bawendi, M. G., Cathodoluminescence and photoluminescence of highly luminescent CdSe/ZnS quantum dot composites. *Appl. Phys. Lett.* 1997, *70,* 2132-2134.

19. García de Abajo, F. J.; Howie, A., Relativistic Electron Energy Loss and Electron-Induced Photon Emission in Inhomogeneous Dielectrics. *Phys. Rev. Lett.* 1998, *80,* 5180.

20. Shuford, K. L.; Ratner, M. A.; Schatz, G. C., Multipolar excitation in triangular nanoprisms. *J. Chem. Phys.* 2005, *123,* 114713.

21. Hao, E.; Schatz, G. C., Electromagnetic fields around silver nanoparticles and dimers. *J. Chem. Phys.* 2004, *120,* 357-366.

22. Martin, O. J. F., Plasmon Resonances in Nanowires with a Non-regular Cross-Section. In *Optical Nanotechnologies*, Tominaga, J.; Tsai, D. P., Eds. Springer-Verlag: Berlin Heidelberg, 2003; pp 183-210.

23. Nelayah, J.; Gu, L.; Sigle, W.; Koch, C. T.; Pastoriza-Santos, I.; Liz-Marzán, L. M.; van Aken, P. A., Direct imaging of surface plasmon resonances on single triangular silver nanoprisms at optical wavelength using low-loss EFTEM imaging. *Opt. Lett.* 2009, *34,* 1003-1005.

24. Rang, M.; Jones, A. C.; Zhou, F.; Li, Z. Y.; Wiley, B. J.; Xia, Y. N.; Raschke, M. B., Optical Near-Field Mapping of Plasmonic Nanoprisms. *Nano Lett.* 2008, *8,* 3357-3363.

25. http://www.lumerical.com/

26. Raether, H., *Surface-Plasmons on Smooth and Rough Surfaces and on Gratings*. 1988; Vol. 111, p 1-133.

27. Garcia de Abajo, F. J.; Kociak, M., Probing the photonic local density of states with electron energy loss spectroscopy. *Phys. Rev. Lett.* 2008, *100,* 106804.

28. Johnson, P. B.; Christy, R. W., Optical constants of noble metals. *Phys. Rev. B* 1972, *6,* 4370-4379.

29. Sherry, L. J.; Jin, R. C.; Mirkin, C. A.; Schatz, G. C.; Van Duyne, R. P., Localized surface plasmon resonance spectroscopy of single silver triangular nanoprisms. *Nano Lett.* 2006, *6,* 2060-2065.

30. Hsu, K. H.; Schultz, P. L.; Ferreira, P. M.; Fang, N. X., Electrochemical Nanoimprinting with Solid-State Superionic Stamps. *Nano Lett.* 2007, *7,* 446-451.

31. Chaturvedi, P.; Hsu, K.; Zhang, S.; Fang, N., New Frontiers of Metamaterials: Design and Fabrication. *MRS Bull.* 2008, *33,* 915-920.

32. Palik, E. D., *Handbook of optical constants*. 1984; Vol. 1.

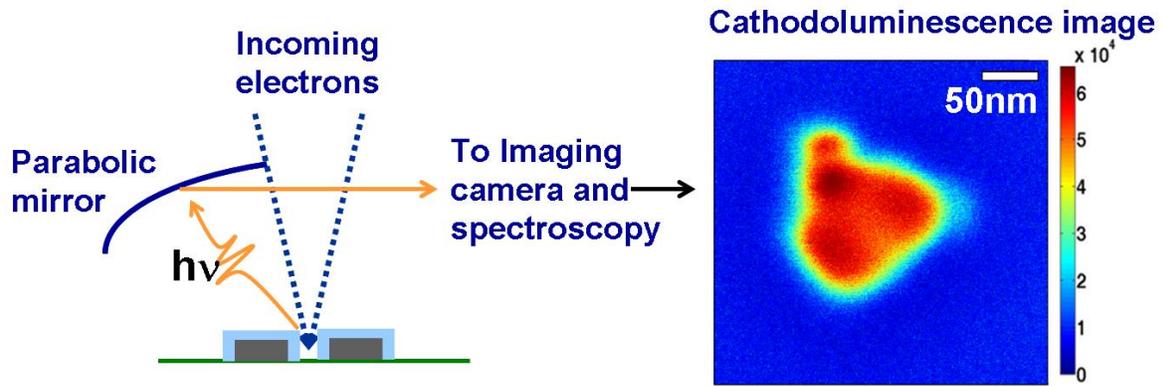

Scanning electron beam based cathodoluminescence spectroscopy allows high-resolution imaging of plasmon modes